\documentclass[]{spie}
\usepackage[dvips]{graphicx}

\title{Background Measurements from Balloon-Borne CZT Detectors} 

\author{
  Johnathan A Jenkins\supit{a}, 
  Tomohiko Narita\supit{b},
  Jonathan E. Grindlay\supit{a},
  Peter F. Bloser\supit{c}, \\
  Carl Stahle\supit{d},
  Brad Parker\supit{d}, and
  Scott Barthelmy\supit{d}
  \skiplinehalf
  \supit{a}Harvard-Smithsonian Center for Astrophysics, 60 Garden Street, 
  Cambridge, MA~02138, USA \\
  \supit{b}College of the Holy Cross, Worcester, MA~01610, USA \\
  \supit{c}Max-Planck-Institute f${\rm \ddot{u}}$r extraterrestische Physik,\\ 
 Giessenbachstrasse, D-85748 Garching, Germany \\
  \supit{d}NASA Goddard Space Flight Center, Greenbelt, MD~20771, USA
}

\authorinfo{Further author information: (Send correspondence to J.A.J.)\\
J.A.J.: E-mail: jjenkins@cfa.harvard.edu}

\begin{document} 
\maketitle 

\begin{abstract}
We report detector characteristics and background measurements 
from two prototype imaging CdZnTe (CZT) detectors flown on a 
scientific balloon payload in May 2001.
The detectors are both platinum-contact 
10 mm $\times$ 10 mm $\times$ 5 mm
CZT crystals, each with a 4 $\times$ 4 array of pixels tiling
the anode.  
One is made from IMARAD horizontal Bridgman CZT, 
the other from eV Products high-pressure Bridgman CZT. 
Both detectors were mounted side-by-side
in a flip-chip configuration 
and read out by a 32-channel IDE VA/TA ASIC preamp/shaper.  
We enclosed the detectors in the same 
40$^\circ$ field-of-view collimator (comprising
a graded passive shield and plastic scintillator) used in
our previously-reported September 2000 flight.  
I-V curves for the detectors are diode-like, and we find that
the platinum contacts adhere significantly better to the
CZT surfaces than gold to previous detectors.  The detectors
and instrumentation performed well in a 20-hour balloon flight
on 23/24 May 2001.  Although we discovered a significant
instrumental background component in flight, it was possible to
measure and subtract this component from the spectra.  The 
resulting IMARAD detector background spectrum (from 30 keV 
to $\sim$ 450 keV) reaches 
$\sim 5 \times 10^{-3}$ counts cm$^{-2}$s$^{-1}$keV$^{-1}$
at 100 keV and has a power-law index of $\sim$ 2 at high energies.
The eV Products detector has a similar spectrum, although there
is more uncertainty in the energy scale because of calibration
complications.
\end{abstract}


\section{INTRODUCTION}

A sensitive survey of the sky in hard X-rays ($\sim$ 10 - 600 keV) will
require a coded-aperture design for imaging over a wide field of view,
and this in turn requires a very large-area detector which is reasonably
efficient even at the upper end of the energy band and has high
resolution in position, energy, and  time (see Grindlay et
al, these Proceedings and astro-ph/0211415, for a description of 
the proposed EXIST mission). The most promising strategy 
for achieving these goals involves a high-Z, room-temperature semiconductor
detector (CdZnTe, hereafter CZT) read out through an array of anode electrode
pixels by multi-channel ASIC electronics.

In section 2 we will briefly describe the CZT2 experiment, emphasizing differences
from the previous flight \cite{Bloser02}. We used entirely new detectors for this
flight, and the preparation and characterization of these detectors (one made from
an eV Products crystal and the other from an IMARAD crystal) is described in section 3.

We had a successful, 20-hour balloon flight with CZT2  (section 4),
which flew alongside several other instruments. In the course of data analysis
(section 5), we discovered a significant instrumental background component
of the spectrum using the multi-channel readout capability of our VA-TA ASIC,
and we show two compatible methods for subtracting this (cosmic-ray
related) instrument background to obtain photon 
background spectra which are relevant to future
wide field-of-view coded-aperture telescopes. We conclude by explaining how our
laboratory and balloon program will progress toward the goal of large-area
survey telescopes by incorporating cathode readouts for depth-sensing and
large-area modular detectors with integrated electronics. 

\section{THE CZT2 EXPERIMENT}

CZT2 consists of two 10 mm $\times$ 10 mm $\times$ 5 mm CZT detectors
flip-chip mounted to a carrier board, a 32-channel VA-TA ASIC readout from IDE AS,
an ASIC controller board and PC-104 computer, shield assembly and readout,
and power supplies and filters for the
electronics. All of the equipment is enclosed in a cylindrical pressure vessel so the
experiment can be flown with one atmosphere of nitrogen to resist high-voltage
breakdown and to facilitate cooling of warm components at balloon altitude.

CZT2 was designed to measure the background of a wide field-of-view pixel-array
CZT detector under high-altitude, near-space conditions. 
A previous planar-detector CZT experiment (CZT1, see Bloser et al. 
1998\cite{Bloser98}, Bloser et al. 2002\cite{Bloser02}) 
explored the effectiveness of an active rear BGO shield on a wide-FOV detector;
CZT2 relies on a graded Pb-Sn-Cu shield and an active plastic scintillator shield
for charged particle rejection. Both of these approaches (BGO and plastic active shielding)
remain in consideration for a survey mission. 

\subsection{Instrumental configuration}

We have described CZT2 in its original form previously \cite{Bloser02}, so here
we emphasize the modifications used for the current flight; aside from the new
platinum-contact detector (section 3), there are three main improvements to
the set-up: (1) better grounding between components of the system and several new
cables and power supplies optimized for lower noise; we were able to operate
CZT2 with a lower discriminator setting throughout the flight, yielding better
low-energy response. (2) Removal of an $^{241}$Am calibration source in the FOV;
the calibration source dominated the sky background in the previous flight,
and nearly eliminated sensitivity below 70 keV. Without the calibration source,
we were able to measure the background down to 30 keV. Furthermore, since we
re-used the ASIC from the previous flight, we re-used the temperature-gain relation
measured by centroiding the $^{241}$Am photopeak to sharpen the spectra
obtained in the absence of an in-flight calibration. (3) Switch from custom
serial-port to ethernet link between the CZT computer and out central EXITE
computer. Ethernet provides a faster a more reliable connection between
pressure vessels, which increases the maximum data rate the system can process,
reduces dead time accrued when the CZT computer is sending data rather
than acquiring it, and simplifies the custom software by making greater use of
off-the-shelf libraries capable of reliably transmitting data structures with simple
procedure calls. Other current and planned instruments in our group (CZT3, EXITE 3/4)
will also use commercial high-speed serial technologies such as ethernet and USB
wherever possible to shift computing power away from the detectors toward
centrally-located flight computers which process data for many detector modules.
Figure 1 shows a schematic diagram of the CZT2 detector assembly and support electronics.

 \begin{figure}
 \begin{center}
   \begin{tabular}{cc}
   \includegraphics[height=5cm]{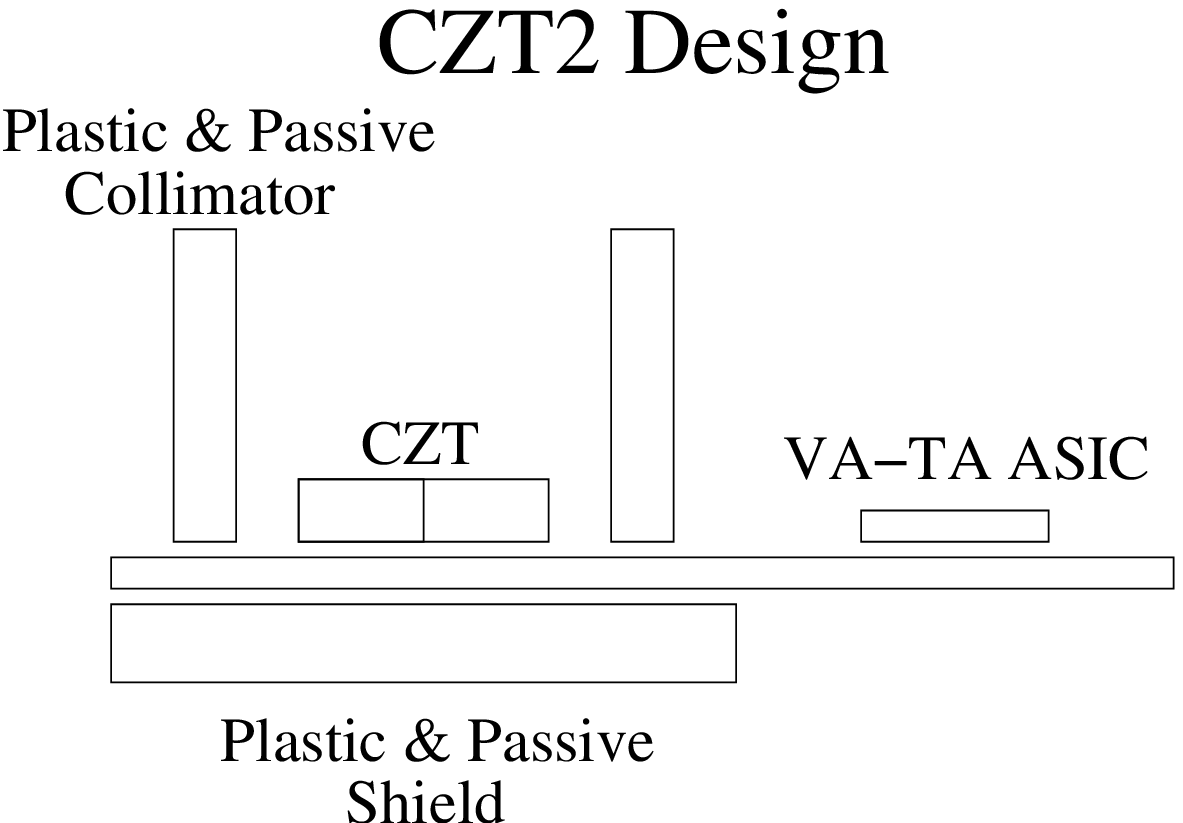} &
   \includegraphics[height=5cm]{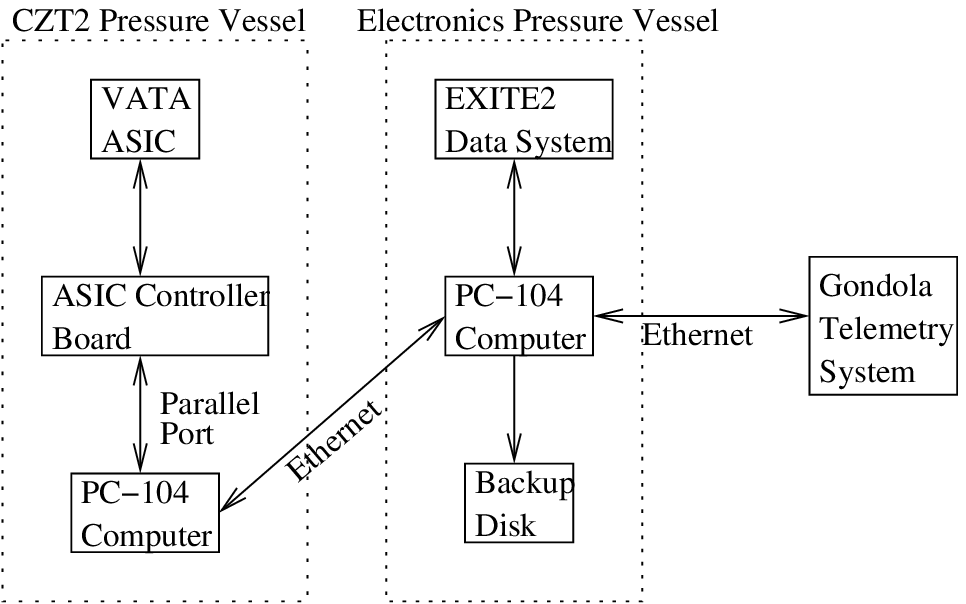}
   \end{tabular}
   \end{center}
   \caption[example] 
   { \label{fig:example} 
Schematic diagram of the CZT2 detector, shield, and readout board assembly (left),
and command/data schematic (right).}
   \end{figure}

\subsection{Previous flight}

On September 19, 2000, CZT2 flew for the first time \cite{Bloser02} with two
gold-contact CZT detectors (one eV material, the other IMARAD) and measured
a background of $\sim 4 \times 10^{-3}$ counts cm$^{-2}$ s$^{-1}$ keV$^{-1}$
at ~100keV. The background spectrum was roughly flat below 100 keV
(although the in-flight calibration source limited sensitivity below 70 keV),
and rolled over into a power-law with photon index $\sim 2$ above 100 keV.

\section{DETECTOR FABRICATION AND CHARACTERISTICS}

\subsection{Detector Material and Contacts}

For the May, 2001 flight we replaced the detectors in CZT2, again using one
high-pressure Bridgeman eV crystal and one horizontal Bridgeman IMARAD
crystal, but this time applying platinum contacts. In our previous work 
with gold-contact IMARAD CZT detectors \cite{Narita00} 
we have found that 
despite forming a Schottky barrier with p-type IMARAD material and thus reducing
dark current through the detector, the gold has variable adhesion properties 
such that a relatively low yield of operable detectors was obtained 
with IMARAD CZT using Au contacts. We substituted platinum 
contacts for gold in this flight in the hope that Pt would preserve the electrical 
properties of Au against CZT but adhere more reliably.

The detectors were prepared at Goddard using the same contact application
procedure for both crystals. Each bare crystal (eV and IMARAD) was first etched
with dilute bromine methanol solution to remove surface damage. Then $\sim 700 \AA$
of platinum was evaporated onto the crystal using a shadow mask. The anode 
electrode pattern, a $4 \times 4$ array of pixels surrounded by a thin guard
ring, is identical to the pattern we used in the September 2000 flight. After passivation,
the surface resistivity is $\sim 1.2 \times 10^{11} {\rm \Omega}$ between the cathode and a
single anode pixel.

\subsection{Dark Current Measurements}

After the flight, we measured the dark current in each pixel as a function of
bias voltage in order to evaluate the resistance properties of the detectors
in detail. 

CZT2 is not optimized for low-noise readout in a number of ways;
e.g., the circuit board connecting the ASIC to the detector socket has leads
which are both long and of variable length, and the computer controlling the
experiment is located in the same pressure vessel as the detector and ASIC.
This means that CZT2 is not capable of pushing energy resolution down to
the limits of the detector/ASIC combination (a few keV) where dark current
contributions to system noise would reveal themselves directly in the FWHM.
The following dark current $I-V$ measurements thus represent a short cut; a
way of evaluating the ultimate noise contribution to the system due to dark
current, even when that noise is too small to appear in spectra.

Figure 2 shows I-V curves for the eV-material detector. Each plot in the grid corresponds
to the physical position of a pixel on the detector anode. The curious result here
is the variation in the curves from pixel to pixel: some pixels have curves which
suggest Schottky barriers for negative bias (e.g. second row pixels),  
others seem ohmic (e.g. bottom row, left two pixels). The ohmic pixel 
behaviour seen with Au-contacts on eV Products CZT\cite{Narita00} was
with significantly lower leakage current than found here.  
We speculate that the eV Products CZT in the present 
detectors may contain inclusions or be non-uniform
in some other way, perhaps with respect to zinc fraction, so that parts of 
the surface are slightly
n-type while others are slightly p-type or intrinsic. When platinum (slightly p-type)
contacts are applied, I-V curves and the presence of diode-like behavior
could then depend on the precise composition of the crystal under a particular
anode pixel.

Figure 3 shows the corresponding plot for IMARAD material, which is slightly n-type.
In this case, we see much more consistent Schottky barrier behavior, with the
exception of the upper-left pixel. Because this is only a single pixel in the
corner of the detector, it may be due to surface leakage cased by imperfect
passivation. Both the IMARAD and eV Products detectors have $\sim 1-2$ nA
dark current at -500 V bias in the best pixels, with all but one of
the IMARAD detector pixels at $<$2 nA vs. only 4 of the eV detector
pixels. 

 \begin{figure}
 \begin{center}
   \begin{tabular}{c}
   \includegraphics[height=10cm]{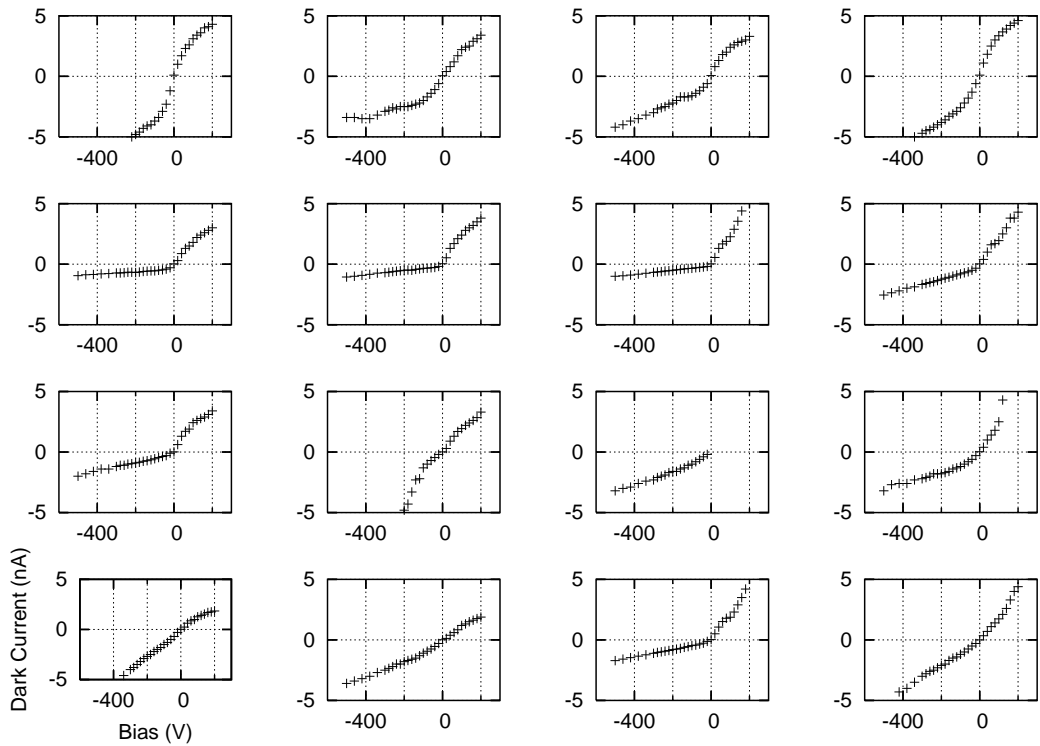}
   \end{tabular}
   \end{center}
   \caption[example] 
   { \label{fig:example} 
I-V dark current curves for each pixel in the eV products CZT detector. Each box
corresponds to the physical position of an anode pixel on the detector surface. The
variation among the curves may be due to material variations.}
   \end{figure} 

 \begin{figure}
  \begin{center}
   \begin{tabular}{c}
   \includegraphics[height=10cm]{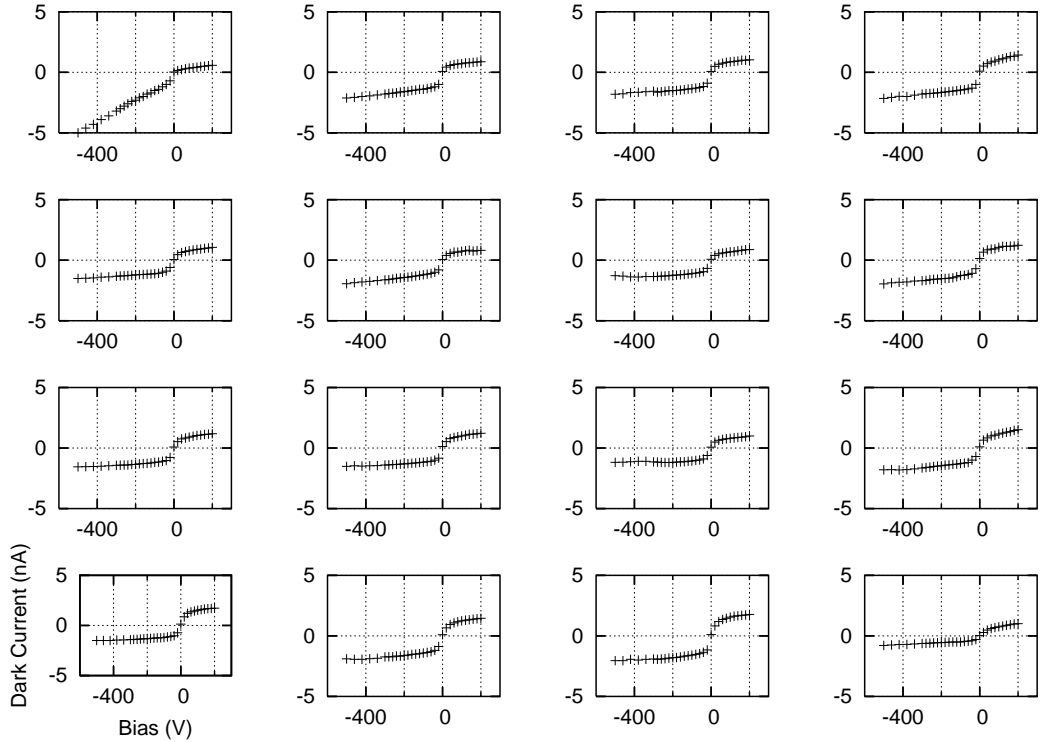}
   \end{tabular}
   \end{center}
   \caption[example] 
   { \label{fig:example} 
I-V curves for the IMARAD CZT detector. The anomalous, quasi-ohmic behavior of
upper-left pixel is likely caused by side leakage in one corner of the detector.}
   \end{figure} 

\section{BALLOON FLIGHT RESULTS}

We had a successful 20-hour balloon flight from Fort Sumner, NM beginning with launch
at 16:24 UTC, 15 May 2001. The flight was a collaboration with the HERO hard x-ray 
focusing optics group at MSFC, and was also the final flight of Harvard's EXITE2 
phoswitch/coded-aperture telescope \cite{Bloser02a}. 
CZT2 acquired data at float altitudes of 119,000 ft to 126,000 ft for $\sim 15$ hours.

\subsection{Energy Calibration and Temperature correction}

We calibrated the ASIC pulse height/energy relation using six spectral lines
from three different radioactive sources ($^{241}$Am, $^{57}$Co, and $^{133}$Ba) 
placed successively 
in the field of view while CZT2 was being tested on the ground. The six lines
(or rather, five lines plus the $^{241}$Am escape peak) range in energy 
from 30 keV to 356 keV, and show that the ASIC is nearly linear in that range,
although saturates slightly above 500 keV (with the exact value depending on
the channel) with our gain settings.

The ASIC channels connected to the eV detector are significantly noisier, on average,
than those connected to the IMARAD detector because of longer circuit board 
traces between the ASIC and detector. Because of this, the calibration data
for the eV detector is incomplete (many eV channels had to be masked out
during the calibration runs).
The IMARAD detector spectra are calibrated, and we concentrate on spectral
results from that detector.

Because the ASIC and its support electronics were unchanged from the September
2000 flight, we assume that the effect of temperature on the system gain and
pedestal offset is
similar to that observed in the earlier flight, with a calibration source, even though
we did not fly a source during the May 2001 flight. 
We correct for temperature
using a simple model derived from Bloser et al. 2002, figure 4, in which the gain
 does not change over the relevant temperature range for the 2001 flight
($15 - 23$ C), and the pedestal
offset (as measured by the location of the in-flight $^{241}$Am cal source 
60 keV peak and the 
ASIC low-energy noise peak) 
increases linearly with temperature.

\subsection{Subtraction of the Instrumental Background}

During test runs on the ground before flight, the system showed very low background
in the absence of test sources. In flight, however, we measured a background
component {\em even  in ASIC channels which were electronically disabled}.
The VA-TA ASIC has a feature which allows us to disable channels so they will
not trigger a readout (viz., photons interacting in the corresponding anode pixels
will be ignored). Nevertheless, disabled channels are still read out when an event
in some other (active) pixel triggers a readout. The spectrum obtained from a
disabled channel represents either a measure of ambient noise
in the system, or some kind of residual glow caused by photons or cosmic rays
triggering active  channels but also depositing charge in the disabled channels.

We have found that high-energy events in the disabled channels tend not to be
correlated with high-energy events in the active channels; this suggests that the
background in the disabled channels is, in fact, an instrumental background.
Because it appears only at high-altitude, it must be cosmic-ray related, perhaps
the result of noise induced by the elevated shield rates or 
charged particles interacting with some stage of the 
readout electronics. Under the assumption that we are seeing an instrumental
background component, it follows that a similar component should be present
in the spectra obtained from the active channels. The photon background spectrum
for a wide FOV detector, dominated by aperture flux but also containing locally-produced
photons, should be obtained by subtracting a disabled-channel spectrum from
an active-channel spectrum to remove the instrumental noise events. We approach
this in two ways, described below.

\subsubsection{Instrumental Background Subtractions in Individual Pixels}

We limited our telemetry rate for the CZT2 experiment to 50 events per second
to reduce dead time caused by computer processing. Given this constraint, there
is a trade-off: we could choose a low lower-level discriminator (LLD) setting and
push the low-energy response down at the expense of having to disable many of
the noisier channels, or we could raise the LLD and activate more channels. 
During the flight, we used two configurations most of the time. 
In the first configuration (cf. spectra in figure 4), 
five channels are disabled (those marked 8 and 17 - 20) 
and the LLD is set to an energy corresponding to
roughly 30 keV. In the second configuration, four of these channels 
(8, 17, 18, and 20) are re-activated
as the LLD is re-set to correspond to $\sim 45$ keV.

 \begin{figure}
 \begin{center}
   \begin{tabular}{c}
   \includegraphics[height=11cm]{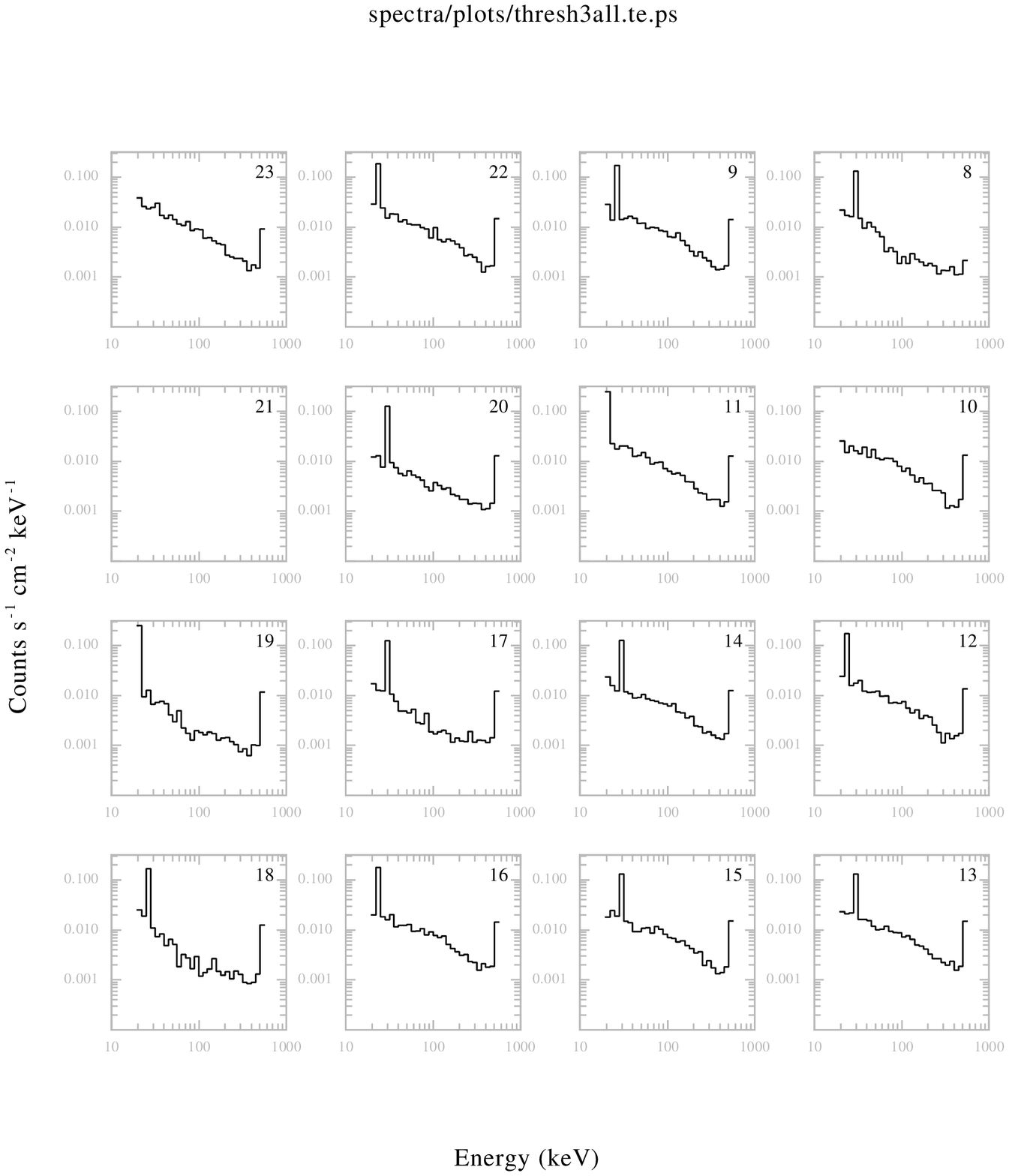}
   \end{tabular}
   \end{center}
   \caption[example] 
   { \label{fig:example} 
IMARAD CZT ``good'' (non-vetoed) event spectra. The channels marked 8 and 17 - 20
are disabled, and reveal a cosmic-ray related instrumental background component.}
   \end{figure} 

As long as we restrict attention to energies above 45 keV, we can subtract
the instrumental background on a channel-by-channel basis in a straightforward
way, but only for the four channels which switched from being disabled to active
as the LLD was raised: we divide each block of data by the appropriate integration
time, and subtract a channel's extracted spectrum in the ``disabled'' data set from
that in the ``active'' data set. We show the results of this procedure in figure 5.
 \begin{figure}
 \begin{center}
   \begin{tabular}{c}
   \includegraphics[height=10cm]{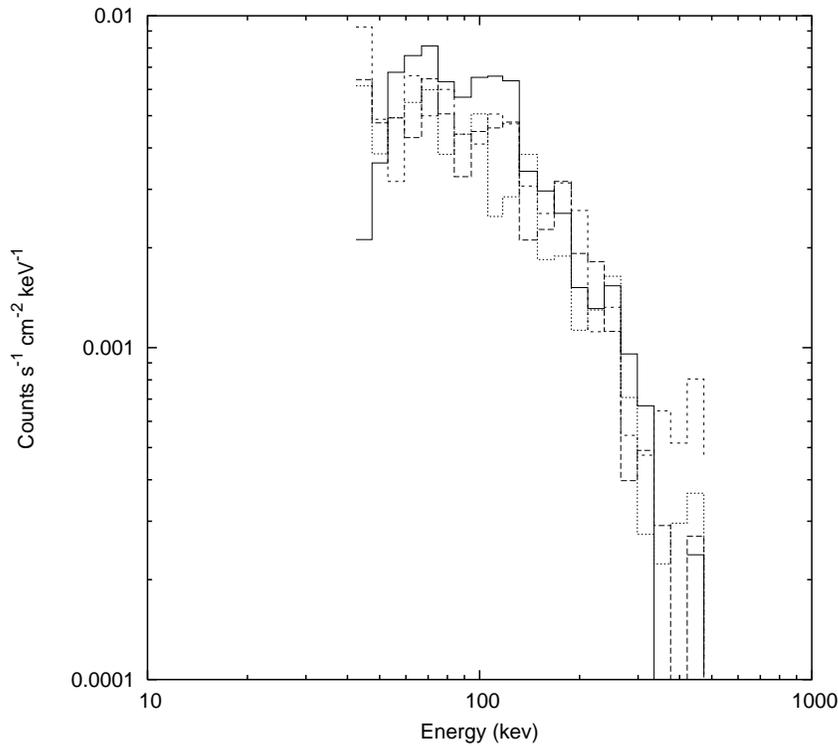}
   \end{tabular}
   \end{center}
   \caption[example] 
   { \label{fig:example} 
CZT2 flight background spectrum from 4 ASIC channels 
(8, 17, 18, and 20) where the instrumental
CR-related background component has been subtracted channel-by-channel.
Each represents a distinct ASIC channel. Good events only (non-vetoed)}
   \end{figure}

\subsubsection{Instrumental Background Subtraction Averaging Many Pixels}

The method described above suffers from two disadvantages. First, the data
come from only 4 of the 15 working IMARAD channels, so relatively few
photons go into each spectrum and statistical uncertainties are high. Second, 
the data is limited to energies higher than 45 keV, the higher of the two 
LLD settings.

An alternative approach is to assume that the gains, efficiencies, and pedestal
offsets of all pixels are similar, and in a single data set (namely, the 30 keV
threshold data) average together the 10 active channel spectra and the
5 disabled-channel spectra, and subtract the latter average from the former. We
show the results of this method, which is clearly vulnerable to systematic
uncertainties resulting from variations among the channels, in figures 6
(average spectra before subtraction) and 7 (subtracted spectum).
 \begin{figure}
 \begin{center}
   \begin{tabular}{c}
   \includegraphics[height=10cm]{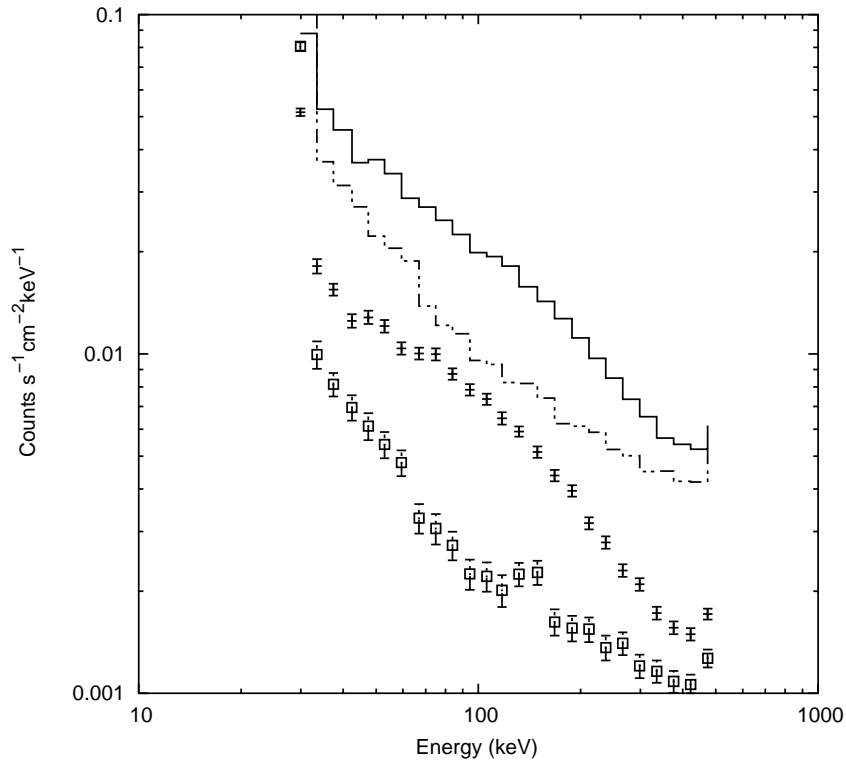}
   \end{tabular}
   \end{center}
   \caption[example] 
   { \label{fig:example} 
CZT2 flight background spectrum showing averaged disabled channels
(instrumental background from channels 8 and 17 - 20) and averaged enabled channels
(total background including the instrumental component). Figure 7 shows
the result of removing the instrumental background component by 
subtracting the lower two curves.
Solid line: averaged spectrum (enabled channels); all events.
Dotted line: averaged spectrum (disabled channels); all events. 
Data points (and errors): averaged spectrum 
(enabled channels); good events only (not vetoed).
Data-squares (and errors): averaged spectrum 
(disabled channels); good events only (not vetoed). The latter two 
are subtracted in figure 7.}
   \end{figure} 

\subsection{Comparison with Predicted Background and Previous Flight}

Our background measurements are consistent with the measurements made
during the 2000 flight, although they extend the low-energy response from 70
down to $\sim 30$ keV. Bloser et al. (2002) describe an MGEANT
simulation of the CZT2 background spectrum which includes
contributions from aperture flux (which is dominant), shield leakage, and
local production due to cosmic ray interactions with the apparatus\cite{Bloser02}. 
Figure 7 shows spectra from the 2001 flight along with the simulation results.
 \begin{figure}
 \begin{center}
   \begin{tabular}{c}
   \includegraphics[height=10cm, angle=-90]{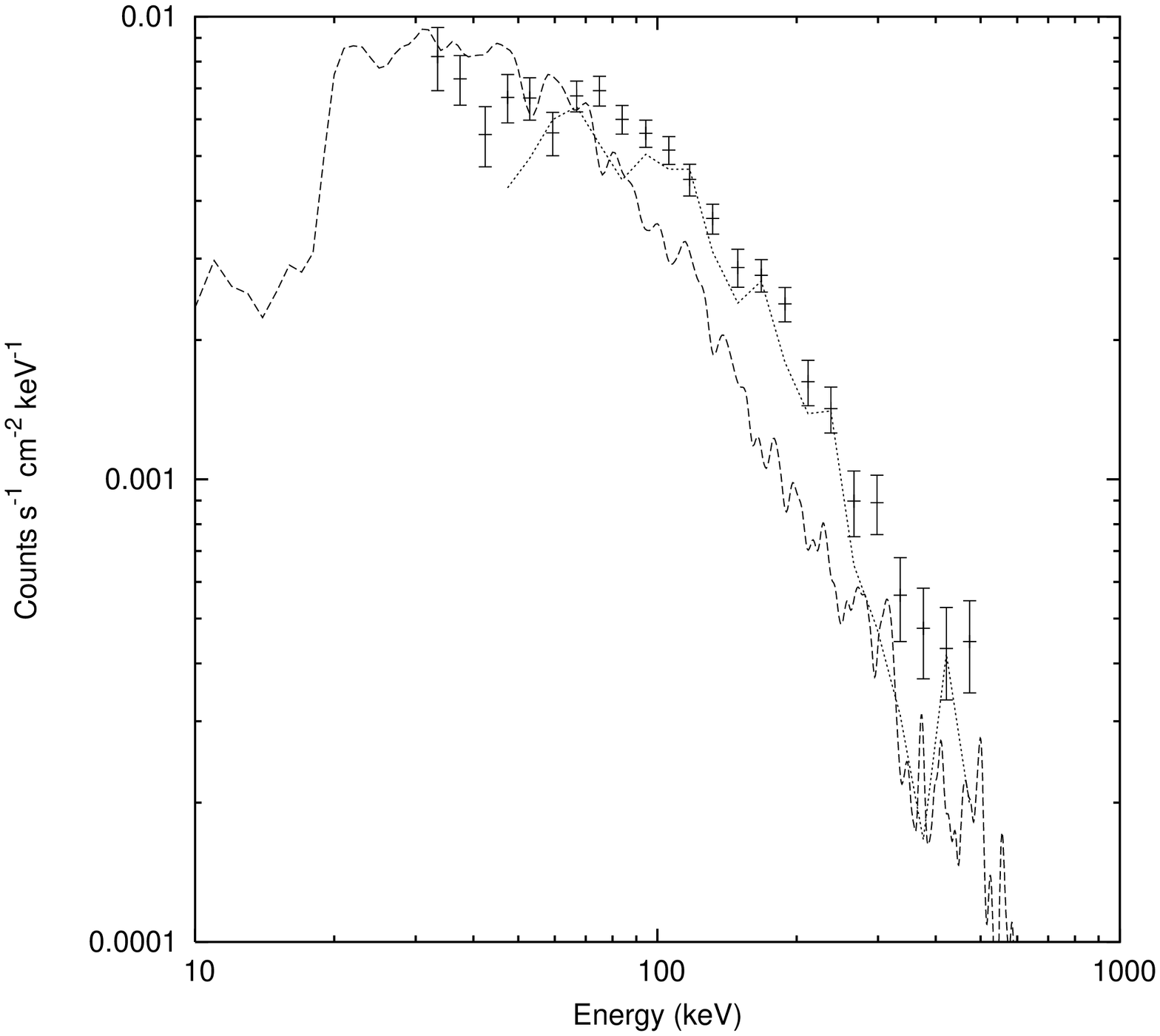}
   \end{tabular}
   \end{center}
   \caption[example] 
   { \label{fig:example} 
Comparison of background hard x-ray spectra from 
the Spring 2001 flight (specifically, the difference of the lower two `non-vetoed' 
curves in figure 6) and MGEANT simulations (dashed line). 
C.f. Bloser '02 
figure 8\cite{Bloser02} for the data from the earlier flight, and the text in that
paper for a detailed description of the simulations. For comparison,
the dotted line represents the average of the individually-subtracted spectra
in figure 5.}
   \end{figure} 

\section{CONCLUSIONS AND FUTURE WORK}

We have found that platinum contacts adhere well to CZT crystal, and
provide reliable current-limiting Schottky barriers with IMARAD material
in the same way gold does. Performance is variable, even over a single
detector, for our detector made from eV Products CZT. Using two new
detectors and other improvements to the system, we had a successful
balloon flight with the CZT2 instrument during which we obtained a measurement
of the background at high altitude for a wide-FOV CZT detector with active
plastic and passive lead-tin-copper graded shields. There was a significant
instrumental background, probably related to cosmic rays interacting with the
electronics, at high altitude; we characterized and subtracted this component
of the background by taking advantage of the multi-pixel readout capability
of the VA-TA ASIC. The resulting subtracted, detector-only background, dominated
by aperture gamma-ray flux but also containing a component caused by local
production of photons, is similar to what we have measured before: a flat spectrum
at $\sim 6 \times 10^{-3}$ counts s$^{-1}$ cm$^{-2}$ kev$^{-1}$ from $\sim$30-100 keV,
turning over into a photon-index 2 power-law at higher energies. This spectrum
is in rough agreement with our MGEANT simulations, although the normalization,
turnover point, and high-energy slope do not line up precisely with what we
measured.

We have constructed a larger-area (8 cm $\times$ 8 cm) CZT detector, CZT3,
which incorporates the IDE XA ASIC in an integrated module with
the detector carrying board\cite{Narita02}, and we have begun to
incorporate a cathode readout into that system (also using IDE ASICs)
to provide  depth-sensing capability. Future balloon
instruments (EXITE3 and EXITE4)
will be large-area scientific detectors (up to 0.5 m$^2$) composed of tiled
CZT/ASIC modules similar to those in CZT3. These instruments will incorporate
cathode readouts and will also test thicker CZT detectors, detectors with
smaller pixel pitch, and different ASIC technologies over the next several years.

\section{ACKNOWLEDGEMENTS}

We thank B. Ramsey, J. Apple, K. Dietz, and the rest of the MSFC HERO
team for technical support on our shared balloon gondola, and NSBF for a 
successful flight. B. Sundal at IDE AS and U. El-Hanany at IMARAD provided
help with the VA-TA ASIC readout and CZT crystals respectively. 
Francis C. Niestemski at the College of the Holy Cross 
made many of the detector I-V measurements.
This work was
supported in part by NASA grants NAG5-5103 and NAG5-5209, and J. A. J.
acknowledges support from an NSF Graduate Student Research Fellowship.

\bibliography{spie} 
\bibliographystyle{spiebib} 

\end{document}